\documentclass[english,prd,twocolumn,superscriptaddress,nofootinbib,preprintnumbers]{revtex4}
\usepackage[latin1]{inputenc}
\usepackage{enumerate}
\usepackage{graphicx}
\usepackage{amssymb}
\usepackage{color}
\usepackage{float}
\usepackage{amsmath}
\usepackage{amsfonts}
\usepackage{dcolumn}
\usepackage{hyperref}
\usepackage{subfigure}
\usepackage{epsfig}
\usepackage{epsf}
\usepackage{epstopdf}
\begin{document}
 
\title{Global structure of Black Holes via dynamical system}
 
\author{Apratim Ganguly}
\email{212559839@stu.ukzn.ac.za}
\affiliation{Astrophysics and Cosmology Research Unit, School of Mathematics, Statistics and Computer Sciences, University of KwaZulu-Natal, Private Bag X54001, Durban 4000, South Africa.}
 
\author{Radouane Gannouji}
\email{radouane.gannouji@ucv.cl}
\affiliation{Instituto de F\'isica, Pontificia Universidad Cat\'olica de Valpara\'iso (PUCV), 23-40025 Valpara\'iso, Chile.}
\affiliation{Astrophysics and Cosmology Research Unit, School of Mathematics, Statistics and Computer Sciences, University of KwaZulu-Natal, Private Bag X54001, Durban 4000, South Africa.}
 
\author{Rituparno Goswami}
\email{Goswami@ukzn.ac.za}
\affiliation{Astrophysics and Cosmology Research Unit, School of Mathematics, Statistics and Computer Sciences, University of KwaZulu-Natal, Private Bag X54001, Durban 4000, South Africa.}
 
\author{Subharthi Ray}
\email{rays@ukzn.ac.za}
\affiliation{Astrophysics and Cosmology Research Unit, School of Mathematics, Statistics and Computer Sciences, University of KwaZulu-Natal, Private Bag X54001, Durban 4000, South Africa.}
 

\begin{abstract}
We recast the system of Einstein field equations for Locally Rotationally Symmetric spacetimes into an autonomous system of covariantly defined geometrical variables. The analysis of this autonomous system gives all the important global features of the maximal extension of these spacetimes. We conclude that the dynamical system analysis can be a powerful mathematical tool for qualitative understanding of the global structure of spacetimes  covariantly, without actually solving the field equations.
\end{abstract}

\maketitle

\section{Introduction}

In general relativity (GR), any spacetime can be regarded as a solution to the Einstein field equations $G_{ab}=T_{ab}$, if one defines the energy momentum tensor of the matter 
according to the left hand side of the equation, that can be calculated from the metric tensor of the spacetime. However the matter tensor so defined will in general have unphysical properties and in most of the cases will have no resemblance to the standard matter around us. Hence by the term {\em exact solution} of Einstein field equations we shall mean the following:  A spacetime (${\cal M},{\bf g}$) in which the field equations are satisfied with the energy momentum tensor ($T_{ab}$) of some specific form of matter which obeys the postulate of local causality and at least one of the physically reasonable energy conditions \cite{Hawking:1973uf}. Most of the well known exact solutions are thus for the empty space ($T_{ab}=0$), for an electromagnetic field, for a perfect fluid or for combination of these. Because of the extreme complexity of the field equations, which are in general 10 coupled non linear second order partial differential equations, it is impossible to find exact solutions except in the spaces of high symmetry (e.g. spherical symmetry) and for relatively simple matter content. In this regard these exact solutions are rather idealised. 

Nevertheless, the exact solutions give the idea of important qualitative features that can arise in GR and hence the possible properties of the realistic solutions of field equations. One of the most intriguing and challenging task is to find the global properties of the field equations by the maximal analytic extension of the local solutions. Study of these global structure of the solutions are important as we get the maximal manifold (${\cal M},{\bf g}$) on which the solution is valid and hence the maximal complete atlas. This enables us to get rid of all the coordinate singularities that may appear due to bad choice of coordinates while solving the field equations. Obtaining such maximal extension may be tedious and tricky as one 
needs to cleverly redefine the spacetime coordinates so that the space around the coordinate singularity  becomes regular. By this step, we get rid of the coordinate singularity and the metric tensor becomes nondegenerate even in the locus of the previous coordinate singularity. We may continue it as far as we can till this process ultimately stops because the spacetime is surrounded either by asymptotic infinity - infinite volume where trajectories may be extended to an infinite proper length - or by genuine (curvature) singularities that cannot be extended by any coordinates. Geodesics physically terminate at those real singularities.

We know dynamical systems approach has proven to be a very important mathematical tool in studying the global properties of various cosmologies in GR \cite{WainEllis} and also other higher order theories of gravity \cite{Carloni:2004kp,Amendola:2006we,Goheer:2008tn,Zhou:2009cy,Xiao:2011nh,Leon:2012mt,Heisenberg:2014kea,Chiba:2014sda,Kofinas:2014aka}. Similar analysis were performed to study the properties of spherically symmetric solutions in dimensionally reduced spacetimes and diatonic black holes in GR and other higher order theories of gravity \cite{Mignemi:1988qc, Wiltshire:1990ah, Mignemi:1991wa, Poletti:1994ff, Mignemi:1999zy, Melis:2005xt, Clifton:2005aj}. The most important advantage of dynamical systems technique is that without solving the system completely one can have qualitative informations on important global features of the phase space, in terms of the fixed points of the system, their stabilities and different invariant submanifolds of the complete phase space.

The aim of this paper is as follows:
\begin{enumerate}[(a)]
\item Using a semitetrad covariant formalism, we show that one can recast the field equations (which are the combination of Ricci and doubly contracted Bianchi identities) for vacuum (with or without a cosmological constant) or electrovacua Locally Rotationally Symmetric (LRS-II) spacetimes into an autonomous system of covariantly defined variables. Hence by definition, this autonomous system is gauge independent.
\item Using the usual Poincar\'e compactification, we compactify the phase space of this autonomous system.
\item Using the general symmetries of LRS-II spacetimes and the properties of the phase space of the above defined autonomous system, we show that we can have the qualitative idea of all the important global features of these spacetimes, without actually solving the system.
\end{enumerate}
Thus the analysis developed in this paper can be effectively used to find the important global properties of other more realistic solutions of Einstein field equations, without solving these equations. 
 
In this paper, we confine our attention to spherically symmetric  vacuum (with or without a cosmological constant ) or electrovacuum, see \cite{standby} for applications to modified gravity theories. For technical reasons it is convenient to consider a class of spacetimes which are a small generalisation of spherically symmetric metrics: namely {\it Locally Rotationally Symmetric} (LRS) class II spacetimes \cite{Ellis:1966ta,Stewart:1967tz,vanElst:1995eg}.  These are evolving and vorticity free spacetimes with a 1-dimensional isotropy group of spatial rotations at every point. Except for  few higher symmetry cases, these spacetimes have locally (at each point) a unique preferred spatial direction that is covariantly defined. 
To describe these spacetimes in terms of metric components, it is well known that the most general interval for LRS-II is written as \cite{Stewart:1967tz}
\begin{eqnarray}
   ds^2 &=& -A^{2}(t,\chi)\,dt^2+B^2(t,\chi)\,d\chi^2 \nonumber\\
   &&+C^2(t,\chi)\,[\,dy^2+D^2(y,k)\,dz^2\,] \;,\label{LRSds}
\end{eqnarray}
where $t$ and $\chi$ are parameters along the integral curves of the timelike vector field  $u^a = A^{-1}\delta^a_0$ and the preferred spacelike vector field $e^a = B^{-1}\delta^a_\nu$. The function $D(y,k)= \sin y,y, \sinh y$ for $k=(1,0,-1)$ respectively. The 2-metric $\,dy^2+D^2(y,k)\,dz^2$ describes spherical, flat, or open homogeneous and isotropic 2-surfaces for $k=(1,0,-1)$. Spherically symmetric spacetimes are the $k=1$ subclass of these spacetimes.  
One can easily see that all the physically interesting spherically symmetric spacetimes fall in the class LRS-II.  

It has been recently shown in \cite{Goswami:2011ft}, that a vacuum or electrovac LRS-II spacetime (with or without a cosmological constant) has an extra symmetry in terms of existence of a Killing vector in local $[u,e]$ plane, where $u^a$ and $e^a$ are timelike and spacelike vector fields respectively, defined above. This extra Killing vector, if timelike, makes the spacetime locally static and if spacelike makes the spacetime localy spatially homogeneous. In the maximally extended manifold these two sections are joined via a 3 dimensional submanifold commonly know as the event horizon. Using this extra symmetry of LRS-II spacetimes, we recast the field equations into a covariantly defined autonomous system separately for both these sections, compactify the phase spaces and show that we can recover all the important features of the global properties of these solutions.

\section{1+1+2 Covariant approach}

The formalism follows the same strategy as the $1+3$ decomposition or threading of space-time (see \cite{Roy:2014lda} for a comparison with the so-called $3+1$ formalism, or slicing of space-time), where one split the spacetime onto a timelike and an orthogonal three-dimensional spacelike hypersurface. All information is captured in a set of kinematic and dynamic variables. We can further decompose the 3-hypersurface into a spacelike vector and a 2-space. This strategy was developed in \cite{Clarkson:2002jz,Clarkson:2007yp} (see also \cite{Maeda1980} for the so-called $2+1+1$ formalism). In this paper we will study the simple problem of spherically symmetric spacetimes, hence the full set of variables are scalars \cite{Clarkson:2002jz} which simplifies the analysis. In fact, the same situation happens in cosmology where the space is homogeneous and isotropic and by virtue of the symmetry the $1+3$ decomposition gives rise to equations evolving only scalars. Therefore the only non-zero variables for any rotationally symmetric spacetime are scalars in the $1+1+2$ approach, therefore it is natural approach for study of LRS spacetimes.

\subsection{Formalism} 

First we perform a standard $1+3$ decomposition. For this, we define a unit timelike vector $u^a$ $(u^a u_a=-1)$ which defines the projection tensor on the 3-space $h^a_b=g^a_b+u^au_b$. Hence we can define two derivatives; one following the vector $u^{a}$ defined as
\begin{align}
\dot{T}^{a..b}{}_{c..d}{} = u^{e} \nabla_{e} {T}^{a..b}{}_{c..d}~, 
\end{align}
and a projected derivative defined as 
\begin{align}
D_{e}T^{a..b}{}_{c..d}{} = h^a{}_f h^p{}_c...h^b{}_g h^q{}_d h^r{}_e \nabla_{r} {T}^{f..g}{}_{p..q}~, 
\end{align}
Further we perform the split of the 3-space by introducing a unit spacelike vector $n^{a}$ 
\begin{align}
n_{a} u^{a} = 0\;,\; \quad n_{a} n^{a} = 1.
\end{align}
with a projection tensor on the 2-space (sheet) orthogonal to $n^a$ and $u^a$
\begin{align} 
N_{a}{}^{b} \equiv h_{a}{}^{b} - n_{a}n^{b} = g_{a}{}^{b} + u_{a}u^{b} 
- n_{a}n^{b}~,~~N^{a}{}_{a} = 2~, 
\label{projT} 
\end{align} 
Hence we can define two additional derivatives along $n^a$ in the surface orthogonal to $u^a$
\begin{align}
\hat{T}_{a..b}{}^{c..d} &\equiv  n^{f}D_{f}T_{a..b}{}^{c..d}~, 
\end{align}
and a projected derivative onto the sheet
\begin{align}
\delta_e T_{a..b}{}^{c..d} &\equiv N_{a}{}^{f}...N_{b}{}^gN_{i}{}^{c}..N_{j}{}^{d}N_e{}^kD_k T_{f..g}{}^{i..j}\,.
\end{align} 

\subsection{Variables} 

The Riemann curvature tensor represents completely the spacetime which is fully determined by the Weyl tensor (free gravitational field) and the Ricci tensor which is determined locally at each point by the energy-momentum tensor. Hence in a fully $1+3$ covariant approach, we split the Weyl curvature tensor $W_{abcd}$ relative to $u^a$ into electric $E_{ab}=W_{a c b d}u^c u^d$ and magnetic $H_{ab}=\frac{1}{2}\epsilon_{a c d}W^{c d}_{~~b e}u^e$ parts and $\epsilon_{abc}$ is the 3-space permutation symbol. Also the energy-momentum tensor $T_{ab}$ can be decomposed relative to $u^a$
\begin{align}
T_{ab}=\rho u_a u_b+p h_{ab}+q_b u_a+q_a u_b+\pi_{ab}
\end{align}
where $\rho$ is the energy density, $p$ isotropic pressure, $q^a$ momentum density (energy flux) and $\pi_{ab}$ trace-free anisotropic pressure (anisotropic stress). For LRS spacetime, only scalars do not vanish after the additional decomposition of space. Hence the only non-zero part of the heat flux and the anisotropic pressure are
\begin{align}
q_a = Q n_a,~\text{and}~\pi_{ab}=\Pi (n_a n_b-\frac{1}{2}N_{ab})
\end{align}
Also the non-zero part of the electric part of Weyl tensor is $E_{ab}=\mathcal{E}(n_a n_b-\frac{1}{2}N_{ab})$ and we will focus on spherically symmetric spacetimes with $n^a$ points along the radial direction. Hence the spacetime is vorticity free (LRS-II) which further constrains the magnetic Weyl curvature ${\cal H}=0$ \cite{Betschart:2004uu}.
The additional non-zero geometrical quantities are respectively the expansion ($\theta = \nabla_a u^a$), shear ($\Sigma = n^a n^b \nabla_{a} u_b$), sheet expansion ($\phi = \delta_a n^a$) and acceleration ($\mathcal{A} = n^a \dot u_a$)

\subsection{Equations} 

The complete set of propagation and/or evolution equations which define these spacetimes, namely LRS class II spacetimes, are :
\subsubsection{Propagation equations:}
\begin{align} 
\hat\phi &= -\frac{1}{2}\phi^2+\left(\frac{1}{3}\theta + \Sigma \right)\left(\frac{2}{3}\theta -\Sigma \right)\nonumber \\
 &-\frac{2}{3}\left(\rho+\Lambda \right)-\frac{1}{2}\Pi-{\cal E}~,
\label{equation31}\\
\hat\Sigma-\frac{2}{3}\hat\theta &=-\frac{3}{2}\phi\Sigma - Q ~,
\label{equation32}\\
\hat{\cal E} -\frac{1}{3}\hat\rho + \frac{1}{2}\hat\Pi &=- \frac{3}{2}\phi\left({\cal E}+\frac{1}{2}\Pi \right)+\left(\frac{1}{2}\Sigma -\frac{1}{3}\theta \right)Q~,
\label{equation33}
\end{align}  
\subsubsection{Evolution equations:}
\begin{align}  
\dot\phi &=-\left(\Sigma -\frac{2}{3}\theta\right)\left({\cal A}-\frac{1}{2}\phi\right)+Q~,
\label{equation34}\\
\dot\Sigma -\frac{2}{3}\dot\theta &= - {\cal A}\phi + 2\left(\frac{1}{3}\theta -\frac{1}{2}\Sigma\right)^2 \nonumber \\
&+\frac{1}{3} \left(\rho +3p -2\Lambda \right) -{\cal E} +\frac{1}{2}\Pi~,
\label{equation35}\\ 
\dot{\cal E}-\frac{1}{3}\dot\rho +\frac{1}{2}\dot\Pi &=\left(\frac{3}{2}\Sigma -\theta\right){\cal E}+\frac{1}{4}\left(\Sigma -\frac{2}{3}\theta\right)\Pi \nonumber \\
&+\frac{1}{2}\phi Q-\frac{1}{2}\left(\rho +p\right)\left(\Sigma -\frac{2}{3}\theta\right)~,
\label{equation36}
\end{align}   
\subsubsection{Mixed (Propagation/Evolution) equations:}
\begin{align} 
\hat{\cal A}-\dot\theta &= -\left({\cal A}+\phi\right){\cal A} + \frac{1}{3}\theta ^2+\frac{3}{2}\Sigma ^2\nonumber \\
&+\frac{1}{2}\left(\rho +3p-2\Lambda \right)~, 
\label{equation37}\\
\dot\rho+\hat Q &= -\theta\left(\rho +p\right)-\left(\phi+2{\cal A}\right)Q-\frac{3}{2}\Sigma\Pi ~,
\label{equation38}\\
\dot Q+\hat p+\hat\Pi &= -\left(\frac{3}{2}\phi+{\cal A}\right)\Pi -\left(\frac{4}{3}\theta+\Sigma\right)Q \nonumber \\
&-\left(\rho+p\right){\cal A}~.
\label{equation39}
\end{align}  
In most general case we will consider only electromagnetic field. Assuming that we do not have magnetic monopole or using the duality rotation \cite{Plebanski:2006sd}, we can always suppress the magnetic field in the vacuum. Also the electric field can be decomposed in the form $E^a=E n^a$ which is solution of $\hat E=-\phi E$ and $\dot E=(\Sigma-\frac{2}{3}\theta)E$. We have $F_{\mu \nu}=\frac{1}{2}u_{[\mu} E_{\nu]}$ from which we have 
\begin{align}
T_{\mu \nu}=\frac{E^2}{\mu_0}\Bigl[\frac{1}{2}g_{\mu \nu}+u_\mu u_\nu-n_\mu n_\nu\Bigr]
\end{align} 
which gives $Q=0$, $\Pi=-4\rho/3$, $P=\rho/3$ and $\rho=E^2/2\mu_0$. We can always absorb the constants and work with the variable $\rho$ which is solution of the equations
\begin{align}
\hat \rho &= -2\phi \rho ~, 	\label{propCharge}\\
\dot \rho &= 2(\Sigma-\frac{2}{3}\theta)\rho ~.	\label{evoCharge}
\end{align}

We also define  the Gaussian curvature via the Ricci tensor on the sheet $^2R_{ab}=K N_{ab}$ which can be written in the form \cite{Betschart:2004uu}
\begin{align}
K = \frac{1}{3}\left(\rho+\Lambda \right)-{\cal E}-\frac{\Pi}{2}+\frac{\phi^{2}}{4}-\left(\frac{1}{3}\theta -\frac{1}{2}\Sigma\right)^2   \label{GaussCurv}
\end{align}
it gives from the previous equations
\begin{align}
\hat K &= -\phi K ~, \label{propGauss}\\
\dot K &= -\left(\frac{2}{3}\theta-\Sigma\right)K ~.	\label{evoGauss}
\end{align}

Notice that eq.(\ref{GaussCurv}) is a constraint because for any surface $K$ is fixed, e.g. in Schwarzschild coordinates we have $K=1/r^2$. This equation will be used to define the dimensionless variables as the Friedmann equation is used in cosmology.

\subsection{Static case}
In this part we will consider spacetime with an additional timelike killing vector. Therefore all the time derivatives are zero, hence it can easily seen from the previous equations that $\theta=\Sigma=Q=0$. As a consequence the variables  $\left\{{\cal A}, \phi, {\cal E}, \rho, \Lambda \right\}$ fully characterize the kinematics. We define the dimensionless geometrical variables in the following way
\begin{eqnarray}
 x_1=-\frac{{\cal E}}{K},&& x_2=\frac{\phi}{2\sqrt{K}},\nonumber\\
 x_3=\frac{{\cal A}}{\sqrt{K}},&& x_4=\frac{\Lambda}{3K},\nonumber\\
 x_5=\frac{\rho}{K}\;.\label{variable1}
 \end{eqnarray}
 
We have from (\ref{equation31})-(\ref{evoGauss}):  

\begin{align}
\label{s1}
x_1' &= x_2(2 x_5-x_1)~,\\
\label{s2}
x_2' &= \frac{x_1}{2}-x_4~, \\
\label{s3}
x_3' &= x_5 -3 x_4-x_3(x_2+x_3)~, \\
\label{s4}
x_4' &= 2 x_2 x_4 ~,\\
\label{s5}
x_5' &= -2 x_2 x_5 ~,\\
\label{sc1}
0 &= x_1-2 x_4-2 x_2 x_3 ~,\\
\label{sc2}
1 &= x_1+x_2^2+x_4+x_5~,
\end{align}
where we have defined the dimensionless spatial derivative $x'={\hat x}/\sqrt{K}$.

\subsection{Non-static case}

In the previous subsection, we discussed the static case. Here we will assume the presence of spacelike killing vector. Hence all space-derivatives will be zero. Therefore for the non-static Universe, $\phi={\cal A}=Q=0$ and the variables $\left\{\theta, \Sigma, {\cal E}, \rho, \Lambda \right\}$ completely characterize the system. Along with the definitions in (\ref{variable1}), we further define two new variables
\begin{eqnarray}
 x_6=\frac{\theta }{3\sqrt{K}},&& x_7=-\frac{\Sigma}{2\sqrt{K}}\;. \label{variable2}
 \end{eqnarray} 
Here the propagation of the variables will be zero and only the evolution terms remain. The system of equations from (\ref{equation31})-(\ref{evoGauss}), turns out to be:

\begin{align}
\label{ns1}
\mathring{x}_1 &= (2x_5-x_1)(x_6+x_7)~, \\
\label{ns4}
\mathring{x}_4 &=2x_4(x_6+x_7)~, \\
\label{ns5}
\mathring{x}_5 &=-2 x_5(x_6+x_7)~,  \\
\label{ns2}
\mathring{x}_6 &=x_7(x_6-2x_7)+x_4-\frac{x_5}{3}~, \\
\label{ns3}
\mathring{x}_7 &=x_7(2x_7-x_6)+\frac{x_5}{3}-\frac{x_1}{2}~, \\
\label{nsc1}
1 &= x_1+x_4+x_5-(x_6+x_7)^2~, \\
\label{nsc2}
0 &= x_1-2 x_4+2 (x_6-2 x_7)(x_6+x_7)~.
\end{align}
where we define the dimensionless temporal derivative $\mathring{x}=\dot x/\sqrt{K}$. 

\section{Vacuum Spacetime}

In this section we will assume vacuum i.e. $\rho=p=\Pi=\Lambda=0$.

\subsection{Static}

Only the variables $x_1$, $x_2$ and $x_3$ are non-zero. We use the last constraint (\ref{sc2}) to reduce the system to 

\begin{align}
\label{eqs41}
x_2' &=x_2 x_3\\
\label{eqs42}
x_3' &=-x_3(x_2+x_3)\\
\label{eqs43}
1&=2x_2x_3+x_2^2
\end{align}

\begin{figure}
\includegraphics[scale=0.63]{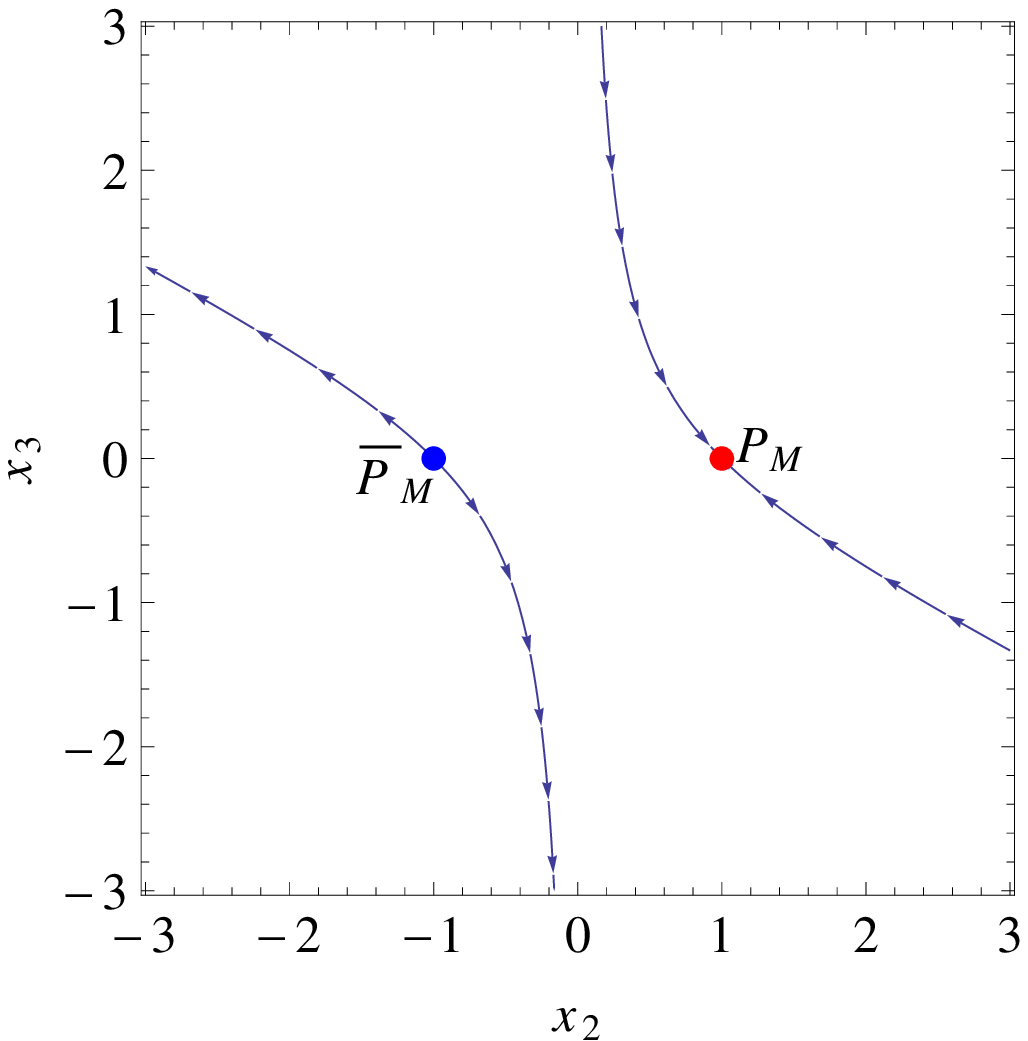}
\includegraphics[scale=0.62]{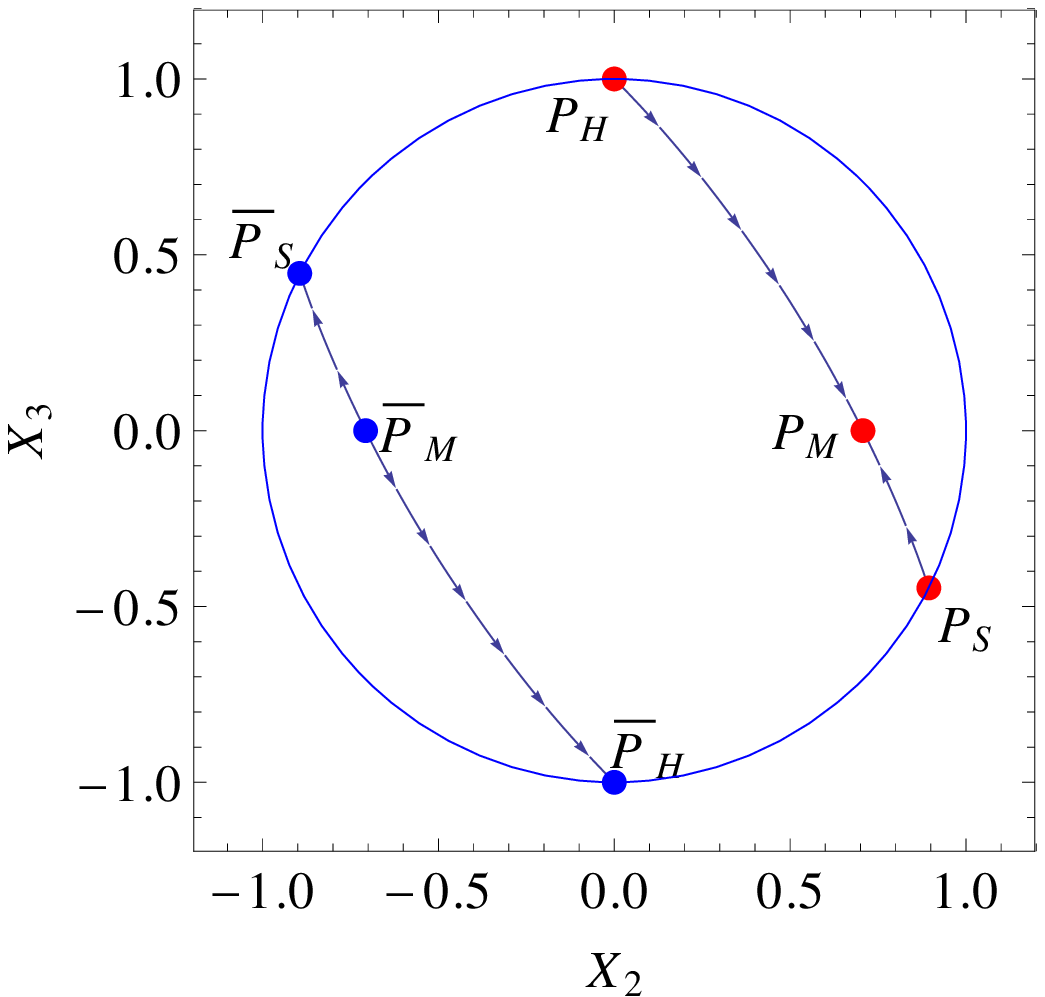}
\caption{The phase portrait for static vacuum in both finite and infinite (Poincar\'e sphere) domains is displayed. The points $(P_M,P_H,P_S)$ corresponds to the black hole solution while $(\bar P_M,\bar P_H,\bar P_S)$ corresponds to the white hole.}
\label{fig:RG}
\end{figure}

The analysis of the system is carried out in the standard way. Notice that the full knowledge of the dynamical system should comprise its behaviour at infinity. Hence we transform the phase space into the so-called Poincar\'e sphere, a sphere with unit radius, tangent to the plane $(x_2,x_3)$ at the origin. Every point of the plane $(x_2,x_3)$ is mapped into 2 points on the surface of the sphere which are situated on the line passing through the point $(x_2,x_3)$ and the center of the sphere. Therefore, infinitely distant points of the plane are mapped into the equator of the sphere. Finally we will represent the orthogonal projection of any one of the hemispheres (to do away with duplicate points) of the sphere onto the tangent plane. This is the projective plane. In the compactified phase portrait, we will use capital letters $(X_2,X_3)$. Under Poincar\'e transformation, the equations become
\begin{align}
X_2'  &= X_2 X_3 (X_2 X_3 + 2 X_3^2 +Z^2)\\
X_3'  &=-X_2^2 X_3 (X_2 + 2 X_3) - X_3 (X_2 + X_3) Z^2\\
Z' &= Z X_3 (-1 + X_2 X_3 + 2 X_3^2 + Z^2)\\
Z^2&=2 X_2 X_3+X_2^2\\
1 &= X_2^2+X_3^2+Z^2
\end{align}
where we have defined $x_i=X_i/Z$ with the constraint $X_2^2+X_3^2+Z^2=1$ and rescaled the derivative $ Z X' \rightarrow X'$. The analysis of the dynamical system for vacuum is summarized in Table \ref{tab:RG} and the phase portrait is shown in Fig.\ref{fig:RG}. Notice that for each point we gave the stability. A hyperbolic equilibrium can be an attractor, repeller or saddle point. But there are many more types for non-hyperbolic equilibria. Most of these equilibria do not have names. A complete classification doesn't exist. Therefore for non-hyperbolic critical points we will specify only if it is stable or unstable.

\begin{table*}[t]
\caption{Critical points and their stability in both finite and infinite (Poincar\'e sphere) domains corresponding to static black hole and white hole.}
\begin{tabular}[c]{l l l l l}
\hline\vspace{-0.4cm}\\
{\centering Dynamical system} & Critical points & Stability & Nature\vspace{0.1cm}\\
  \hline\vspace{-0.3cm}\\
$x_2' =x_2 x_3$ & $P_M:$ $(x_2,x_3)=(1,0)$ & Attractor & Minkowski\\ 
$x_3' =-x_3(x_2+x_3)$ & $\bar P_M:$ $(x_2,x_3)=(-1,0)$ & Repeller & Minkowski\\
$2x_2x_3+x_2^2=1$ & & & \vspace{0.1cm}\\
    \hline\vspace{-0.3cm}\\
$X_2' = X_2 X_3 (X_2 X_3 + 2 X_3^2 +Z^2)$ & $P_H:$ $(X_2,X_3)=(0,1)$ & Repeller & Horizon\\ 
$X_3' =-X_2^2 X_3 (X_2 + 2 X_3) - X_3 (X_2 + X_3) Z^2$ & $\bar P_H:$ $(X_2,X_3)=(0,- 1)$ & Attractor &Horizon\\ 
$X_2^2+X_3^2+Z^2=1$ & $P_S:$ $(X_2,X_3)=(\frac{2}{\sqrt{5}},-\frac{1}{\sqrt{5}})$ & Repeller & Singularity\\
$X_2^2+2 X_2 X_3=Z^2$ & $\bar P_S:$ $(X_2,X_3)=(-\frac{2}{\sqrt{5}},\frac{1}{\sqrt{5}})$ & Attractor & Singularity\\
\hline \hline
\end{tabular}
\label{tab:RG}
\end{table*}
Finally we need to find the nature of each critical point. There are various ways to do it. It can be derived by solving the linearised equations around the critical points. First, we need to define a coordinate system. We will use the spherical coordinates with a metric in the following form
\begin{align}
\label{metric}
{d}s^2=-A{d}t^2+\frac{{d} r^2}{B}+r^2{\rm}d\Omega^2
\end{align}
In order to understand the nature of critical point, we need to determine the different variables in terms of the metric. From the definition (\ref{metric}), we define the four-velocity $u^t=1/\sqrt{A}$ and the radial vector $n^r=\sqrt{B}$. Hence we get
\begin{align}
\label{def:A}
{\cal A} &\equiv -u^a u^b \nabla_a n_b =\frac{\sqrt{B}}{2A}\frac{{\rm d}A}{{\rm d}r}\\
\label{def:phi}
\phi &=N^{~a}_b\nabla_a n^b=\frac{2}{r}\sqrt{B}
\end{align}
Also we can rewrite the derivatives in the following form 
\begin{align}
x'=\frac{\hat x}{\sqrt{K}}=r\hat x=r \frac{r\phi}{2}\frac{{\rm d} x}{{\rm d} r}=x_2  \frac{{\rm d} x}{{\rm d} \ln r}\,,\label{deriv}
\end{align}
where $K=1/r^2$ and we have used $\hat x=n^\mu D_\mu x =\sqrt{B}\frac{{\rm d} x}{{\rm d} r}= r\frac{\phi}{2} \frac{{\rm d} x}{{\rm d} r}$. The last equality comes from (\ref{def:phi}). Hence we have
\begin{align}
B=x_2^2\,,\qquad\frac{d\ln A}{d\ln r}=2\frac{x_3}{x_2}\,,\label{metricX}
\end{align}
from which we can easily recover the metric at each critical point. Notice that each critical point is at a fixed value of radial distance $r$, hence it is necessary to perform a linearisation around the point in order to do an integration and recover the gravitational potential $A$.  For example, the solution of the dynamical system reduces around the critical point $(1,0)$ to
\begin{align}
\label{sol:x2}
x_2 & \simeq 1+\frac{\epsilon}{r}\\
\label{sol:x3}
x_3 & \simeq -\frac{\epsilon}{r}, ~~\text{with} ~~\epsilon \ll 1
\end{align}
From (\ref{metricX}) we get 
\begin{align}
A=B=1\pm \frac{2\epsilon}{r}
\end{align}
The point $P_M$ corresponds to the limit where $\epsilon \rightarrow 0$, therefore $P_M$ is the Minkowski spacetime. Notice from (\ref{def:phi}) that $x_2=\sqrt{B}$, hence we have $x_2>0$. But if we take the inner normal to the surface $n^r=-\sqrt{B}$ we will have $x_2=-\sqrt{B}<0$. Hence the phase space $x_2<0$ will be opposite to the subspace $x_2>0$, the nature of the points will be reversed, e.g. an attractor will be repeller (because of sign change in derivative (\ref{deriv})). Also we see from (\ref{def:A},\ref{def:phi}) that reversing the direction of $u^\mu$ has no effect, because of the static nature of the spacetime. We also notice that in the Kruskal-Szekeres coordinates, we have $U^2-V^2=C^{\text{st}}$ when $r=C^{\text{st}}$. Therefore the normal vector to this hypersurface is $n^\mu=(U,-V,0,0)$ or $n^\mu=(-U,V,0,0)$. In these coordinates, inner/outer direction of the spacelike normal vector $n$ corresponds to the transformation $(U\rightarrow -U,V\rightarrow -V)$ which is equivalent to the transformation from exterior region to parallel exterior region. Therefore the phase space corresponding to $x_2<0$ is the parallel exterior region. The analysis covers the static part of the black hole and the white hole.

We can do the same for the points at infinity, e.g. the point $P_S$. In this case, we have $(X_2,X_3,Z)=(\frac{2}{\sqrt{5}},-\frac{1}{\sqrt{5}},0)$. The solution of the dynamical system around this point (the dynamical system is given in the Table \ref{tab:RG}) is
\begin{align}
X_2 &=\frac{2}{\sqrt{5}}\\
X_3 &=-\frac{1}{\sqrt{5}}\\
Z &=\epsilon \sqrt{r},~~\text{with}~~\epsilon \ll 1
\end{align}
which gives
\begin{align}
x_2 &\equiv \frac{X_2}{Z}\simeq \frac{2}{\epsilon\sqrt{5r}}\\
x_3 &\equiv\frac{X_3}{Z}\simeq -\frac{1}{\epsilon\sqrt{5r}}\\
\end{align}
after redefinition of time (constant is absorbed for $A$), we have
\begin{align}
A=B=\frac{4}{5\epsilon^2r}
\end{align}
Therefore in the limit $\epsilon \rightarrow 0$, we conclude that $P_S$ corresponds to the singularity at $r=0$.

Finally $P_H$ is a little bit more subtle. In fact we can't linearise the equations around this point. We notice that in a stationary spacetime, the apparent horizon coincides with the event horizon and the apparent horizon is a marginally trapped surface on which the outgoing null geodesics have zero expansion \cite{Hawking:1973uf}. We define 2 spacelike vectors $(a^\mu,b^\mu)$ on the 2-surface, which define an orthonormal basis with $n^\mu$ the normal spacelike vector to the 2-surface and $u^\mu$ the timelike vector. Hence we can write the metric as
\begin{align}
g_{\mu \nu}=-u_\mu u_\nu + n_\mu n_\nu  +a_\mu a_\nu  + b_\mu b_\nu\,. 
\end{align}
Also the expansion of the outgoing null geodesics is \cite{Hawking:1973uf,Sasaki1980}
\begin{align}
\label{eq:theta}
\Theta=\frac{1}{2}\nabla_\mu k_\nu (a^\mu a^\nu+b^\mu b^\nu )=\frac{1}{2}\nabla_\mu k_\nu N^{\mu \nu}
\end{align}
where $k^\mu=u^\mu+n^\mu$ is the outgoing null vector. Hence (\ref{eq:theta}) can be written as
\begin{align}
\Theta=\frac{1}{2} \Bigl(N^{\mu \nu}K_{\mu \nu}+\delta_\mu n^\mu\Bigr)
\end{align}
where $K_{\mu \nu}=h_\mu^{~\alpha} h_\nu ^{~\beta} \nabla_\alpha u_\beta$ is the extrinsic curvature. Using the decomposition of the extrinsic curvature and the definition of sheet expansion, we have
\begin{align}
\Theta=\frac{1}{2} \Bigl( \frac{2}{3}\theta-\Sigma+\phi \Bigr)=\sqrt{K}\Bigl(x_2+x_6+x_7\Bigr)
\end{align}
Therefore we conclude that $ x_2+x_6+x_7=0$ \cite{Hamid:2014kza} for the apparent horizon ($\Theta=0$) and hence $x_2=0$ for static case, which implies $P_H$ is horizon. 

Hence we see from Fig.(\ref{fig:RG}) that if the system starts from the horizon $(P_H)$ it goes asymptotically to the Minkowski spacetime $(P_M)$ which corresponds to the standard Schwarzschild black hole solution with a positive mass and if the system starts from the singularity $(P_S)$ it evolves also till Minkowski spacetime but without crossing horizon. That solution corresponds to a naked singularity where the mass is negative. The transformation to the extended spacetime $(U\rightarrow -U,V\rightarrow -V)$ is equivalent to $(x_2\rightarrow-x_2,x_3\rightarrow-x_3)$ which gives the other part of the phase space where $\phi<0$ which means anti-gravity or defocusing of geodesics.	

\subsection{Non-static}

For this case, only the variables $x_1$, $x_6$ and $x_7$ are non-zero. We also use the constraint (\ref{nsc2}) to reduce the system to
\begin{align}
\label{eqns41}
\mathring x_6 &=x_7(x_6-2x_7)\\
\label{eqns42}
\mathring x_7 &=x_6(x_6-2x_7)\\
\label{eqns43}
1&=3\Bigl(x_7^2-x_6^2\Bigr)
\end{align}
There are no finite fixed points. Under Poincar\'e transformation, the equations become
\begin{align}
\mathring X_6 &= -X_7(X_6 - 2 X_7)  (X_6^2 - X_7^2 - Z^2)\\
\mathring X_7 &= X_6 (X_6 - 2 X_7) (X_6^2 - X_7^2 + Z^2)\\
\mathring Z & = -2 X_6 X_7 Z (X_6- 2 X_7)\\
Z^2&=3\Bigl(X_7^2-X_6^2\Bigr)\\
1 &= X_6^2+X_7^2+Z^2
\end{align}
where we have rescaled the derivative $Z \mathring X \rightarrow \mathring X$

We perform the same analysis as before except that the metric takes the following form
\begin{align}
ds^2=-\frac{dt^2}{B(t)}+A(t)dr^2+t^2 d\Omega^2
\end{align}
We define the normal vectors as $u^\mu=(\pm \sqrt{B},0,0,0)$ and $n^\mu=(0,\pm 1/\sqrt{A},0,0)$. It is easy to see that for any field $X$, we have $\dot X=u^\mu \nabla_\mu X=\pm \sqrt{B} dX/dt$. We can also get $\theta=\nabla_\mu u^\mu=(\frac{d \ln A}{dt}/2+2/t)u^0$ and $\theta/3-\Sigma/2=N^{ab}\nabla_a u_b/2=u^0/t$ which gives $u^0=x_6+x_7$. Hence $u^\mu=(x_6+x_7,0,0,0)$. The position of the horizon corresponds to $x_6+x_7=0$. Also the line of constant time are in the Kruskal coordinates defined by $V^2-U^2=C^{st}$ so in these coordinates we have $u^\mu=(V,-U,0,0)$ or $u^\mu=(-V,U,0,0)$. The transformation from non static black hole to non static white hole is ($U\rightarrow -U$, $V\rightarrow -V$) or equivalently by reversing the sign of $x_6+x_7$. As previously the metric can be written in terms of the normalized variables
\begin{align}
B =(x_6+x_7)^2\,,\qquad \frac{d\ln A}{d \ln t} &=2 \frac{x_6-2x_7}{x_6+x_7}\,,
\end{align}
and the derivative $\mathring x=(x_6+x_7)dx/d \ln t$

\begin{figure}[h]
\includegraphics[scale=0.62]{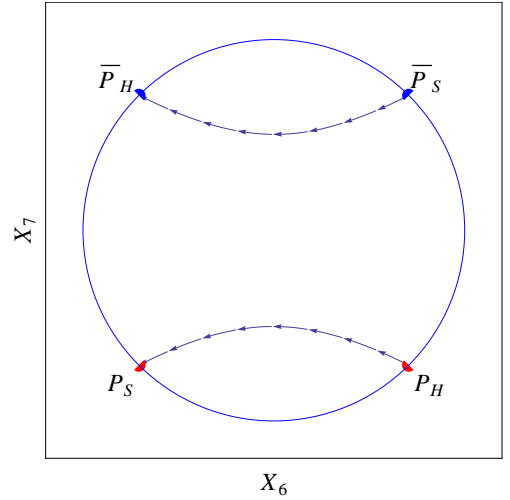}
\caption{The phase portrait for non-static vacuum within Poincar\'e sphere for black hole and white hole.}
\label{fig:RGI}
\end{figure}

Hence it is easy to analyse the system and find the nature of each critical points. The final result is summarized in the Table \ref{tab:IRG} and the phase portrait is shown on Fig.\ref{fig:RGI}.  
\begin{table*}[t]
\caption{Critical points, stability and their nature in both finite and infinite (Poincar\'e sphere) domains for non-static vacuum.}
\begin{tabular}[c]{l l l l}
\hline\vspace{-0.4cm}\\
{\centering Dynamical system} & Critical points & Stability & Nature\vspace{0.1cm}\\
  \hline\vspace{-0.3cm}\\
$\mathring x_6 =x_7(x_6-2x_7)$ & & & \\ 
$\mathring x_7 =x_6(x_6-2x_7)$ & No fixed points & & \\
$1=3\Bigl(x_7^2-x_6^2\Bigr)$ & & & \vspace{0.1cm}\\
    \hline\vspace{-0.3cm}\\
$\mathring X_6 = -X_7(X_6 - 2 X_7)  (X_6^2 - X_7^2 - Z^2)$ & $P_H:$ $(X_6,X_7)=(\frac{1}{\sqrt{2}},-\frac{1}{\sqrt{2}})$ & Repeller & Horizon\\
$\mathring X_7 = X_6 (X_6 - 2 X_7) (X_6^2 - X_7^2 + Z^2)$ & $\bar P_H:$ $(X_6,X_7)=(- \frac{1}{\sqrt{2}},\frac{1}{\sqrt{2}})$ & Attractor & Horizon\\
$Z^2 = 3\Bigl(X_7^2-X_6^2\Bigr)$ &  $P_S:$ $(X_6,X_7)=( -\frac{1}{\sqrt{2}},-\frac{1}{\sqrt{2}})$ & Attractor & Singularity\\ 
$1 = X_6^2+X_7^2+Z^2$ & $\bar P_S:$ $(X_6,X_7)=(\frac{1}{\sqrt{2}},\frac{1}{\sqrt{2}})$ & Repeller & Singularity \\
\hline \hline
\end{tabular}
\label{tab:IRG}
\end{table*}

\section{Vacuum spacetime with cosmological constant}

We follow the same analysis done previously in the presence of a cosmological constant. In this case, $\rho=p=\Pi=0$, so we include $x_4$ other than the variables defined in the previous section. 

\subsection{static}

Using the constraints (\ref{sc1},\ref{sc2}) the system reduces to
\begin{align}
x_2' &=x_2 x_3\\
x_3' &=-x_3(x_3-x_2)+x_{2}^2-1
\end{align}
We see that $x_2 = 0$ is an invariant submanifold of the dynamical system contrary to $x_3=0$, meaning the system can not go through the subspace $x_2 = 0$ and can only approach it asymptotically, which corresponds to horizon as seen previously. We have as before the Minkowski critical point $(x_2,x_3)=(\pm 1,0)$. Using the transformation $x_i=X_i/Z$ with $X_2^2+X_3^2+Z^2=1$, the Poincar\'e sphere, we have
\begin{align}
X_2' &= -X_2 X_3 (X_2^2 + X_2 X_3 - 2 X_3^2 - 2 Z^2)\\
X_3' &= X_2^2 (X_2 - X_3) (X_2 + 2 X_3)\nonumber\\
&\qquad + (X_2- X_3) X_3 Z^2 - Z^4\\
1 &= X_2^2+X_3^2+Z^2
\end{align}
where we performed a rescaling of the derivative $Z X' \rightarrow X'$.

As usual the nature can be derived by making a linearisation around the critical point. Let us consider the point $(\frac{1}{\sqrt{2}},\frac{1}{\sqrt{2}})$. The linearisation gives 
\begin{align}
X_2&\simeq \frac{1}{\sqrt{2}}+\frac{\epsilon_1}{r^3}\\
X_3&\simeq \frac{1}{\sqrt{2}}-\frac{\epsilon_1}{r^3}\\
Z&\simeq \frac{\epsilon_2}{r}
\end{align}
which gives
\begin{align}
x_2 &\simeq \frac{r}{\sqrt{2}\epsilon_2}+\frac{\epsilon_1}{r^2}\\
x_3 &\simeq \frac{r}{\sqrt{2}\epsilon_2}-\frac{\epsilon_1}{r^2}
\end{align}
which at the leading order gives 
\begin{align}
B&\simeq \frac{r^2}{2\epsilon_2^2}\simeq -\frac{\Lambda}{3}r^2\,,\quad \Lambda<0\\
A&\simeq \alpha r^2
\end{align}
where $\alpha$ is a constant of integration and we used the constraint (\ref{sc1},\ref{sc2}) to get $x_4=\Lambda r^2/3\simeq -r^2/2\epsilon_2^2$. Hence we conclude the point is the anti-de-Sitter Universe.

\begin{figure}
\includegraphics[scale=0.62]{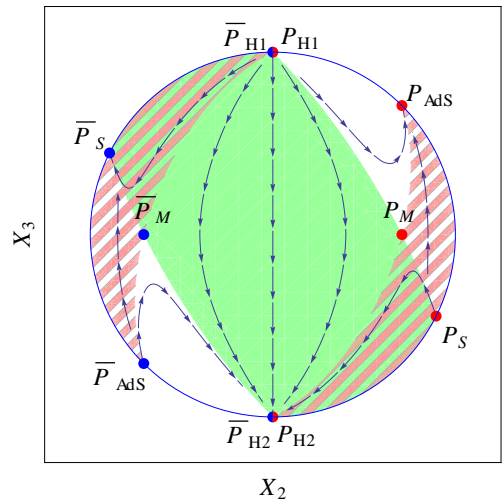}
\caption{The phase portrait for general relativity with cosmological constant within Poincar\'e sphere. The green region corresponds to positive cosmological constant while the dashed red part corresponds to positive electric part of the Weyl tensor ${\cal E} > 0$}
\label{fig:RGL}
\end{figure}

Finally $(P_{H1},\bar P_{H1},P_{H2},\bar P_{H2})$ are horizons. First we notice that because we want a static universe, the sign of the metric can't flip, hence $A>0$ and $B>0$. Also from (\ref{metricX}) we have $\text{sign}(dA/dr)=\text{sign}(x_2 x_3)$. Finally following standard convention, we have $dA/dr>0$ for event horizon and the cosmological horizon is the null surface for which $dA/dr<0$ (or also a cauchy horizon). We conclude that $P_{H1}$ and $\bar P_{H2}$ are event horizons while de-Sitter horizons for $\bar P_{H1}$ and $P_{H2}$. The results are summarized in Table \ref{tab:RGL}. To avoid the singularity, we see from Fig. \ref{fig:RGL} that the sign of the cosmological constant is not important but we avoid it by imposing ${\cal E}<0$.

\begin{table*}[t]
\caption{Critical points and their stability in both finite and infinite (Poincar\'e sphere) domains for general relativity with cosmological constant (static case).}
\begin{tabular}[c]{l l l l}
\hline\vspace{-0.4cm}\\
{\centering Dynamical system} & Critical points & Stability & Nature\vspace{0.1cm}\\
  \hline\vspace{-0.3cm}\\
$ x_2' =x_2 x_3$ & $P_M:$ $(x_2,x_3)=(1,0)$ &Saddle point & Minkowski\\ 
$x_3' =-x_3(x_3-x_2)+x_2^2-1$ & $\bar P_M:$ $(x_2,x_3)=(- 1,0)$ &Saddle point & Minkowski
\vspace{0.1cm}\\
    \hline\vspace{-0.3cm}\\
& $(P_{H1},\bar P_{H1}):$ $(X_2,X_3)=(0,1)$ & Repeller & Horizon\\ 
$X_2'  = -X_2 X_3 (X_2^2 + X_2 X_3 - 2 X_3^2 - 2 Z^2)$ & $(P_{H2},\bar P_{H2}):$ $(X_2,X_3)=(0,-1)$ & Attractor & Horizon\\ 
$X_3' =  X_2^2 (X_2 - X_3) (X_2 + 2 X_3)$ & $P_S:$ $(X_2,X_3)=(\frac{2}{\sqrt{5}},-\frac{1}{\sqrt{5}})$ & Repeller & Singularity\\
\, \, \, \, \, \, \, \, \, $+ (X_2- X_3) X_3 Z^2 - Z^4 $ & & &\\
$Z' =Z X_3 (-2 X_2^2 - X_2 X_3 + X_3^2 + Z^2) $ & $\bar P_S:$ $(X_2,X_3)=(-\frac{2}{\sqrt{5}},\frac{1}{\sqrt{5}})$ & Attractor & Singularity\\
$X_2^2+X_3^2+Z^2=1$ & $P_{AdS}:$ $(X_2,X_3)=(\frac{1}{\sqrt{2}},\frac{1}{\sqrt{2}})$ & Attractor & Anti-de-Sitter\\
& $\bar P_{AdS}:$ $(X_2,X_3)=(-\frac{1}{\sqrt{2}},-\frac{1}{\sqrt{2}})$ & Repeller & Anti-de-Sitter \\
\hline \hline
\end{tabular}
\label{tab:RGL}
\end{table*}
Fig. \ref{fig:RGL} represents the complete static manifold. We can e.g. start an evolution from the event horizon of the BH $(P_{H1})$. Depending on the initial conditions, we can choose a path towards the anti-de Sitter space or evolve to the de Sitter horizon in $P_{H2}$. Localized at the cosmological horizon we can imagine a coordinate transformation which is going to smooth that coordinate singularity, but we don't get rid off the central singularity. That transformation is going to reverse $x_2$ and $x_3$ hence we will be at the point $\bar P_{H1}$ which corresponds to the cosmological horizon where now $\phi<0$, hence we are in the other side of the extension of the spacetime. The system will evolve till the point $\bar P_{H2}$ which corresponds to the event horizon. Now close to that horizon, we can use an other transformation which is going to transform the system into $P_{H1}$ which is again the event horizon, we can proceed to the same thing again and again which shows the infinite structure of the complete manifold. Notice also that we have a straight trajectory from $P_{H1}$ to $P_{H2}$, this corresponds to both horizons indistinguishable, it's the degenerate solution. In fact we have in that case $\phi=0$ and from the equations we have $K=\Lambda$ and 
\begin{align}
\frac{B}{2A}\frac{\rm d^2A}{\rm dr^2}+\frac{1}{4A}\frac{\rm d B}{\rm dr}\frac{\rm dA}{\rm dr}-\frac{B}{4A^2}\Bigl(\frac{\rm dA}{\rm dr}\Bigr)^2+\Lambda=0
\end{align}
Notice that this equation is equivalent to $R_{\mu\nu}=\Lambda g_{\mu\nu}$ with a metric given by ${d}s^2=-A{d}t^2+d r^2/B+d\Omega^2/\Lambda$. 
In case where $A=B$, we have $A=\alpha+\beta r -\Lambda r^2$ corresponding to Nariai spacetime. It is interesting to notice from Fig.\ref{fig:RGL} how easily we deduce the absence of singularity for Nariai spacetime and anti-de Sitter asymptotic region.

\subsection{Non-static}

In this part, we investigate the non-static case with a cosmological constant. Using the constraints (\ref{nsc1},\ref{nsc2}) the system reduces to
\begin{align}
\mathring x_6 &=x_6 x_7-3{x_7}^2+{x_6}^2+\frac{1}{3}\\
\mathring x_7 &={x_7}^2-2 x_6 x_7-\frac{1}{3}
\end{align}
We see that $x_6+x_7$ is an invariant submanifold and hence can't be crossed. Horizons are always invariant submanifolds in our formalism. We have 2 fixed points at finite distance corresponding to horizons. Under Poincar\'e transformation, the equations become
\begin{align}
\mathring X_6 &= 3 X_7^2(X_6^2 - X_7^2) + (3 X_6^2 + 4 X_6 X_7 - 8 X_7^2) \frac{Z^2}{3} + \frac{Z^4}{3}\nonumber \\
\mathring X_7 &= -3 X_6 X_7 (X_6^2 - X_7^2) - (X_6^2 + 7 X_6 X_7 - 3 X_7^2) \frac{Z^2}{3} - \frac{Z^4}{3}\nonumber \\
1 &= X_6^2+X_7^2+Z^2
\end{align}
where we rescaled the derivative ($Z \mathring X \rightarrow \mathring X$).
\begin{table*}[t]
\caption{Critical points and their stability in both finite and infinite (Poincar\'e sphere) domains for general relativity with cosmological constant (non-static case).}
\begin{tabular}[c]{l l l l}
\hline\vspace{-0.4cm}\\
{\centering Dynamical system} & Critical points & Stability & Nature\vspace{0.1cm}\\
  \hline\vspace{-0.3cm}\\
$ \mathring x_6 =x_6 x_7-3{x_7}^2+{x_6}^2+\frac{1}{3}$ & $(P_{H3},\bar P_{H3}):$ $(x_6,x_7)=(\frac{1}{3},-\frac{1}{3})$ &Saddle point & Horizon\\ 
$\mathring x_7 ={x_7}^2-2 x_6 x_7-\frac{1}{3}$ & $(P_{H2},\bar P_{H2}):$ $(x_6,x_7)=(-\frac{1}{3},\frac{1}{3})$ &Saddle point & Horizon\\ 
\vspace{0.1cm}\\
 \hline\vspace{-0.3cm}\\
$ $ & $P_{dS}:$ $(X_6,X_7)=(-1,0)$ & Repeller & de Sitter\\
${\scriptstyle \mathring X_6 = 3 X_7^2(X_6^2 - X_7^2)+ (3 X_6^2 + 4 X_6 X_7 - 8 X_7^2) \frac{Z^2}{3} + \frac{Z^4}{3}}$ & $\bar P_{dS}:$ $(X_6,X_7)=(1,0)$ & Attractor & de Sitter\\ 
${\scriptstyle \mathring X_7 = -3 X_6 X_7 (X_6^2 - X_7^2)- (X_6^2 + 7 X_6 X_7 - 3 X_7^2) \frac{Z^2}{3} - \frac{Z^4}{3}}$ & $P_S:$ $(X_6,X_7)=(-\frac{1}{\sqrt{2}},-\frac{1}{\sqrt{2}})$ & Attractor & Singularity\\
${\scriptstyle \mathring Z = -(X_6^3 + X_6^2 X_7 - 5 X_6 X_7^2 + X_7^3) Z+ (-X_6 + X_7) \frac{Z^3}{3}}$ & $\bar P_S:$ $(X_6,X_7)=(\frac{1}{\sqrt{2}},\frac{1}{\sqrt{2}})$ & Repeller & Singularity\\ 
${\scriptstyle 1 = X_6^2+X_7^2+Z^2}$ & $(P_{H1},\bar P_{H1}):$ $(X_6,X_7)=(-\frac{1}{\sqrt{2}},\frac{1}{\sqrt{2}})$ & Attractor & Horizon\\
 & $(P_{H4},\bar P_{H4}):$ $(X_6,X_7)=(\frac{1}{\sqrt{2}},-\frac{1}{\sqrt{2}})$ & Repeller & Horizon\\ 
 \hline\hline
\end{tabular}
\label{tab:IRGL}
\end{table*}
\begin{figure}
\includegraphics[scale=0.62]{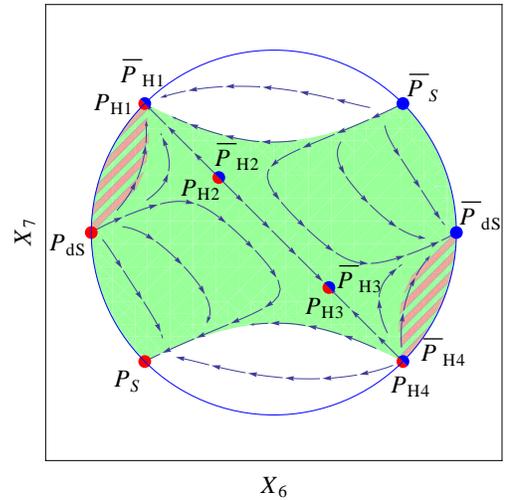}
\caption{The phase portrait for non static with $\Lambda$ in infinite (Poincar\'e sphere) domain. The green part corresponds to $\Lambda>0$ and the red dashed region corresponds to $\mathcal E>0$.}
\label{fig:RGIL}
\end{figure}
The analysis follows the same previous strategy and is summarized in Table \ref{tab:IRGL} and phase space is displayed in Fig. \ref{fig:RGIL}. We notice from Fig. \ref{fig:RGIL} condition ${\cal E}>0$ is sufficient to avoid singularity. Finally the degenerate case where $x_6+x_7=0$ reduces to
\begin{align}
K &= \Lambda \\
\dot \theta+\theta^2-\Lambda &= 0
\end{align}
which gives in terms of metric
\begin{align}
\frac{B}{2A}\frac{\rm d^2 A}{\rm dt^2}-\frac{B}{4A^2}\Bigl(\frac{\rm d A}{\rm d t}\Bigr)^2+\frac{1}{4A}\frac{\rm d A}{\rm d t}\frac{\rm d B}{\rm d t}-\Lambda=0
\end{align}
In the case where $B=\Lambda$ we have $A=\alpha \cosh (t+\beta)^{2}$ corresponding to Nariai solution in global coordinates. We see from the Fig. \ref{fig:RGIL} the solution is singularity-free and do not have asymptotic de Sitter region as expected \cite{Bousso:2002fq}.

\section{Charged Spacetime}
In this section we will consider the presence of a charge hence the additional variable $x_5$. 
\subsection{Static}
\begin{table*}[t]
\caption{Critical points and their stability in both finite and infinite (Poincar\'e sphere) domains for general relativity with charge and cosmological constant (static case).}
\begin{tabular}[c]{l l l l}
\hline\vspace{-0.4cm}\\
{\centering Dynamical system} & Critical points & Stability & Nature\vspace{0.1cm}\\
  \hline\vspace{-0.3cm}\\
 & $(P_{H1},\bar P_{H1}):(x_2,x_3,x_4)=(0,x_3,\frac{1-x_3^2}{6})$ &Saddle line (if $x_3\neq 0)$ & Horizon\\ 
Finite distance & $P_M:(x_2,x_3,x_4)=(1,0,0)$ &Saddle point & Minkowski\\ 
 & $\bar P_M:(x_2,x_3,x_4)=(-1,0,0)$ &Saddle point & Minkowski\\ 
\vspace{0.1cm}\\
    \hline\vspace{-0.3cm}\\
 & $(P_{H2},\bar P_{H2}):(X_2,X_3,X_4)=(0,1,0)$ & Repeller & Horizon\\   
 & $(P_{H3},\bar P_{H3}):(X_2,X_3,X_4)=(0,-1,0)$ & Attractor & Horizon\\
Points at infinity & $(P_{H4},\bar P_{H4}):(X_2,X_3,X_4)=(0,0,1)$ & Stable ($X_2>0$), Unstable ($X_2<0$) & Horizon\\
 & $(P_{AdS},\bar P_{AdS}):(X_2,X_3,X_4)=(0,0,- 1)$ & Stable ($X_2>0$), Unstable ($X_2<0$) & Anti-de Sitter\\
 & $P_{S1}:(X_2,X_3,X_4)=(\frac{2}{\sqrt{5}},-\frac{1}{\sqrt{5}},0)$ & Saddle point & Singularity $(\sim 1/r)$\\
 & $\bar P_{S1}:(X_2,X_3,X_4)=(-\frac{2}{\sqrt{5}},\frac{1}{\sqrt{5}},0)$ & Saddle point & Singularity $(\sim 1/r)$\\
 & $P_{S2}:(X_2,X_3,X_4)=(\frac{1}{\sqrt{2}},- \frac{1}{\sqrt{2}},0)$ & Repeller & Singularity $(\sim 1/r^2)$\\
 & $\bar P_{S2}:(X_2,X_3,X_4)=(- \frac{1}{\sqrt{2}}, \frac{1}{\sqrt{2}},0)$ & Attractor & Singularity $(\sim 1/r^2)$\\
 \hline \hline
\end{tabular}
\label{tab:fullE}
\end{table*}

Using the constraints (\ref{sc1},\ref{sc2}) the equations reduce to 3-dimensional autonomous system
\begin{align}
x_2' &= x_2 x_3\\
x_3' &= 1-3x_2 x_3-x_2^2-x_3^2-6x_4\\
x_4' &= 2 x_2 x_4
\end{align}
with the constraint (positivity of density $\rho=E^2/2\mu_0$)
\begin{align}
x_5=1-x_2^2-3x_4-2x_2 x_3\geq 0
\end{align}
We see that $x_4=0$ and $x_2=0$ are invariant submanifolds. The latter defines the horizon while $x_4 \propto \Lambda$ do not change sign. The critical points and their nature are summarized in Table \ref{tab:fullE}. We have 2 type of singularities which are calculated by a linearisation around the critical point. The weakest singularity $(B\sim 1/r)$ is always a saddle point if $x_5\neq 0$ while the strongest singularity $(B\sim 1/r^2)$ is a repeller for black hole $(x_2>0)$ and an attractor for the white hole. Notice also that $(X_2,X_3,X_4)=(0,0,-1)$ is not a horizon in fact $X_2=0$ doesn't imply $x_2$ zero. This critical point corresponds to the end point of the saddle line in  Table \ref{tab:fullE}. It is stable for the black hole ($X_2>0$) and unstable for the white hole.
\begin{figure}
\includegraphics[scale=0.62]{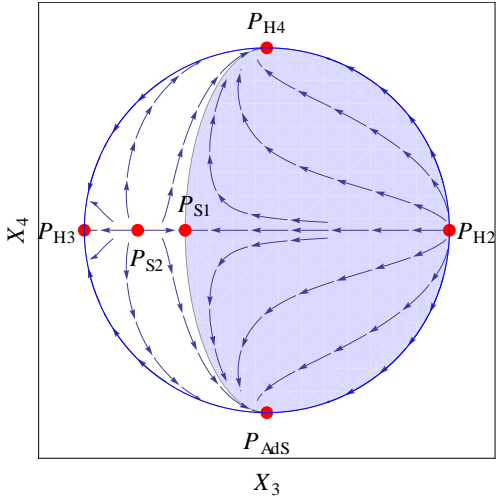}
\caption{The phase portrait at infinity for general relativity with charge and cosmological constant. Only the black hole region is shown $X_2>0$. The white hole phase space can be easily deduced. The blue region represents violation of energy condition $x_5<0$ which is equivalent to $q^2<0$ where $q$ is black hole charge.}
\label{fig:allExt}
\end{figure}
\begin{figure}
\includegraphics[scale=0.62]{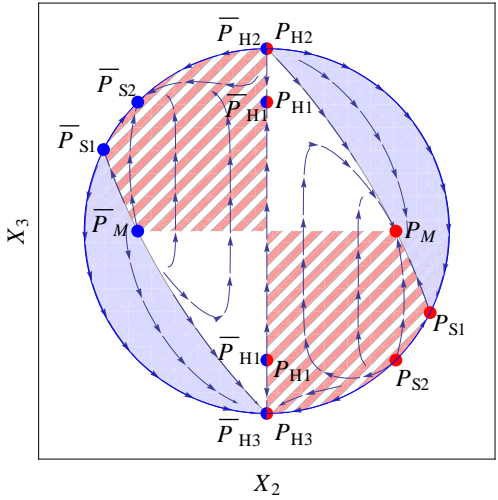}
\caption{The phase portrait for general relativity with charge without cosmological constant in infinite (Poincar\'e sphere) domain. The blue region represents $x_5<0$ and should not be included. The dashed red part represents positive electric part of Weyl tensor ${\cal E}>0$.}
\label{fig:chargeExt}
\end{figure}
In Fig.\ref{fig:allExt}, we have the behaviour of the dynamical system at infinity and Fig.\ref{fig:chargeExt} shows the full phase space for a spacetime without cosmological constant, this is the Reissner-Nordstr\"{o}m solution.

Finally the critical solution $x_2=0$ gives $x_3'+x_3^2+1=0$ which corresponds to Nariai solution. More generically if we impose $\phi=0$ to the equations  (\ref{equation31})-(\ref{evoGauss}) in the static case, and assuming $\Pi=0$, we have $Q=0,p=-\rho,K=\Lambda$ and
\begin{align}
\label{eq:MJ}
\hat {\cal A}+{\cal A}^2+\Lambda =0
\end{align}
It can be integrated easily by defining an affine parameter $\xi$ by $\sqrt{B}d\xi=dr$ which gives ${\cal A}=-\sqrt{\Lambda} \tan(\sqrt{\Lambda}\xi+\alpha)=2^{-1}d\ln A/d\xi$ and hence we have the line element
\begin{align}
{\rm d}s^2 = -\cos^2(\xi) dt^2+\frac{d\xi^2+d\Omega^2}{\Lambda}\,,\quad 
\end{align}
Therefore we can define the static Nariai solution as spacetime without sheet expansion $\phi=0$.

\section{Non-static}

\begin{table*}[t]
\caption{Critical points and their stability in both finite and infinite (Poincar\'e sphere) domains for general relativity with charge and cosmological constant (non-static case).}
\begin{tabular}[c]{l l l l}
\hline\vspace{-0.4cm}\\
{\centering Dynamical system} & Critical points & Stability & Nature\vspace{0.1cm}\\
  \hline\vspace{-0.3cm}\\
& \\
Finite distance & $(P_{H1},\bar P_{H1}):(x_4,x_6,x_7)=(\frac{1+9 x_6^2}{6},x_6,-x_6)$ &Saddle line (if $x_6\neq 0$) & Horizon\\ 
\vspace{0.1cm}\\
    \hline\vspace{-0.3cm}\\
 \\
Points at infinity & $(P_{dS},\bar P_{dS}):(X_4,X_6,X_7)=(1,0,0)$ &  Unstable $(X_6+X_7<0)$ & de Sitter \\
 & $(P_{H2},\bar P_{H2}):(X_4,X_6,X_7)=(-1,0,0)$  &  Unstable $(X_6+X_7<0)$ & Horizon\\
 & $(P_{H3},\bar P_{H3}):(X_4,X_6,X_7)=(0,\frac{1}{\sqrt{2}},-\frac{1}{\sqrt{2}})$  & Repeller & Horizon\\
 & $(P_{H4},\bar P_{H4}):(X_4,X_6,X_7)=(0,- \frac{1}{\sqrt{2}},\frac{1}{\sqrt{2}})$ & Attractor & Horizon \\
 & $\bar P_{S1}:(X_4,X_6,X_7)=(0,\frac{1}{\sqrt{2}},\frac{1}{\sqrt{2}})$  & Saddle point & Singularity $(\sim 1/t)$\\
 & $P_{S1}:(X_4,X_6,X_7)=(0,- \frac{1}{\sqrt{2}},- \frac{1}{\sqrt{2}})$ & Saddle point & Singularity $(\sim 1/t)$\\
 & $\bar P_{S2}:(X_4,X_6,X_7)=(0,\frac{1}{\sqrt{5}},\frac{2}{\sqrt{5}})$ & Repeller & Singularity  $(\sim 1/t^2)$\\
 & $P_{S2}:(X_4,X_6,X_7)=(0,-\frac{1}{\sqrt{5}},-\frac{2}{\sqrt{5}})$  & Attractor & Singularity $(\sim 1/t^2)$\\
 \hline \hline
\end{tabular}
\label{tab:AM}
\end{table*}

Using the constraints (\ref{nsc1},\ref{nsc2}) the non static system reduces to
\begin{align}
\mathring x_4 &= 2 x_4 (x_6+x_7)\\
\mathring x_6 &=x_6 x_7-x_6^2-x_7^2+2x_4-\frac{1}{3}\\
\mathring x_7 &=2x_6(x_6-x_7)-x_7^2-2x_4+\frac{1}{3}
\end{align}
with the constraint on the positivity of density
\begin{align}
x_5=1-3x_4+3x_6^2-3x_7^2 \geq 0
\end{align}
As expected $x_4=0$ is invariant submanifold but also $x_6+x_7=0$ which defines the horizon. The full analysis of the dynamical system is summarized in Table \ref{tab:AM}. We have 2 types of singularities but as in static case $B\sim 1/t$ is a saddle point. The point $(1,0,0)$ corresponds to de Sitter ($\Lambda>0$), it stable for the white hole and unstable for black hole $X_6+X_7<0$.
\begin{figure}
\includegraphics[scale=0.62]{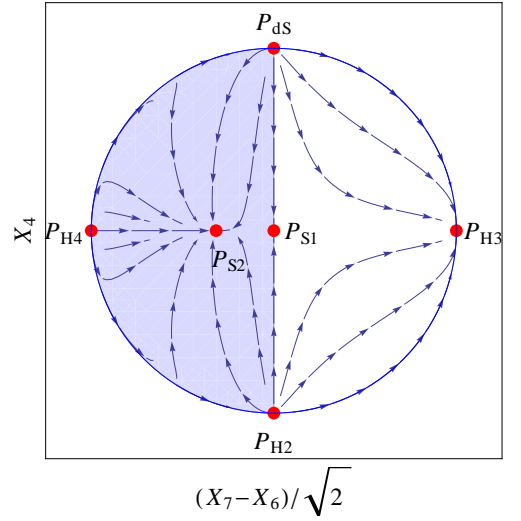}
\caption{The phase portrait at infinity for general relativity with charge and cosmological constant in non-static case. Only one side of the extended manifold is shown. The white region is $\rho>0$.}
\label{fig:allInt}
\end{figure}

\begin{figure}
\includegraphics[scale=0.62]{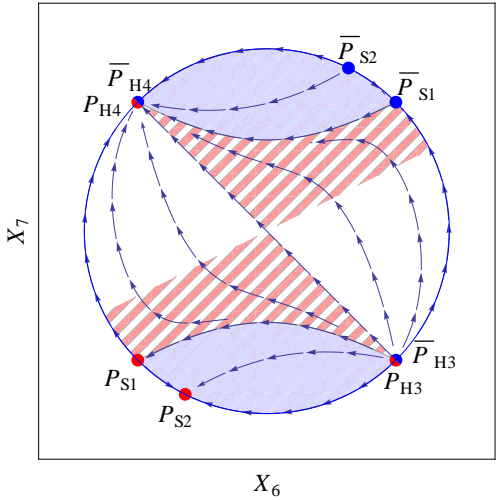}
\caption{The phase portrait for general relativity with charge without cosmological constant in infinite (Poincar\'e sphere) domain. The dashed red region is ${\cal E}>0$ while blue represents forbidden region $x_5<0$.}
\label{fig:chargeInt}
\end{figure}
The behaviour of the full system at infinity is shown in Fig.\ref{fig:allInt} while Fig.\ref{fig:chargeInt} shows the full phase space for $\Lambda=0$. We see that we can't reach the singularity $P_{S2}$ where the metric goes like $1/t^2$ as soon as we assume $\rho=E^2/2\mu_0>0$. In fact to reach the singularity we need to cross an other horizon (cauchy horizon) therefore the spacetime becomes static around this singularity.

Finally, very generically assuming $x_6+x_7=0$ gives from equations  (\ref{equation31})-(\ref{evoGauss}) (and assuming $\Pi=0$)
\begin{align}
K&=\Lambda\\
\dot \theta+\theta^2-\Lambda &=0
\end{align}
As previously, by introducing an affine parameter, it is easy to integrate the equation, we found
\begin{align}
{\rm d}s^2=-\frac{dt^2}{\Lambda}+\cosh^2(t){\rm d}r^2+\frac{{\rm d}\Omega^2}{\Lambda}
\end{align}
Hence imposing the condition $\Sigma = 2\theta/3$ $(x_6+x_7=0)$ for a non-static spherically symmetric spacetime gives Nariai solution.

\section{Conclusion}
In this paper we effectively reformulated the system of Einstein field equations, for LRS-II spacetimes, into an autonomous system of dimensionless, covariantly defined geometrical variables. By compactifying the phase space of this system and using the usual tools of dynamical system analysis we qualitatively found all the important global features of the maximal extension of these spacetimes. Through the construction of this autonomous system of covariant variables we eliminated the problems of coordinate singularities. It is quite interesting that horizons manifest themselves as invariant submanifold of the phase space of the autonomous system. It is also very easy, via this formalism, to see the singularity-free nature of the Nariai solution.

This analysis provides an efficient way to understand the global properties of any spacetime, by bypassing the very difficult task of solving the field equations and maximally extending the solution.

\section{Acknowledgments}

The authors want to thank the National Research Foundation and the University of KwaZulu-Natal for financial support.  R. Gannouji thanks Giacaman.


\begin{thebibliography}{99}

\bibitem{Hawking:1973uf}
  S.~W.~Hawking and G.~F.~R.~Ellis,
  Cambridge University Press, Cambridge, 1973

\bibitem{WainEllis} {\em Dynamical System in Cosmology} edited by Wainwright J and Ellis G F R (Cambridge:
Cambridge Univ. Press 1997) and references therein.

\bibitem{Carloni:2004kp}
  S.~Carloni, P.~K.~S.~Dunsby, S.~Capozziello and A.~Troisi,
  Class.\ Quant.\ Grav.\  {\bf 22} (2005) 4839
  [gr-qc/0410046].

\bibitem{Amendola:2006we}
  L.~Amendola, R.~Gannouji, D.~Polarski and S.~Tsujikawa,
  Phys.\ Rev.\ D {\bf 75} (2007) 083504
  [gr-qc/0612180].

\bibitem{Goheer:2008tn}
  N.~Goheer, R.~Goswami and P.~K.~S.~Dunsby,
  Class.\ Quant.\ Grav.\  {\bf 26} (2009) 105003
  [arXiv:0809.5247 [gr-qc]].

\bibitem{Zhou:2009cy}
  S.~-Y.~Zhou, E.~J.~Copeland and P.~M.~Saffin,
  JCAP {\bf 0907} (2009) 009
  [arXiv:0903.4610 [gr-qc]].

\bibitem{Xiao:2011nh}
  K.~Xiao and J.~-Y.~Zhu,
  Phys.\ Rev.\ D {\bf 83} (2011) 083501
  [arXiv:1102.2695 [gr-qc]].

\bibitem{Leon:2012mt}
  G.~Leon and E.~N.~Saridakis,
  JCAP {\bf 1303} (2013) 025
  [arXiv:1211.3088 [astro-ph.CO]].

\bibitem{Heisenberg:2014kea}
  L.~Heisenberg, R.~Kimura and K.~Yamamoto,
  Phys.\ Rev.\ D {\bf 89} (2014) 103008
  [arXiv:1403.2049 [hep-th]].

\bibitem{Chiba:2014sda}
  T.~Chiba, A.~De Felice and S.~Tsujikawa,
  Phys.\ Rev.\ D {\bf 90} (2014) 023516
  [arXiv:1403.7604 [gr-qc]].

\bibitem{Kofinas:2014aka}
  G.~Kofinas, G.~Leon and E.~N.~Saridakis,
  arXiv:1404.7100 [gr-qc].

\bibitem{Mignemi:1988qc}
  S.~Mignemi and D.~L.~Wiltshire,
  Class.\ Quant.\ Grav.\  {\bf 6} (1989) 987.

\bibitem{Wiltshire:1990ah}
  D.~L.~Wiltshire,
  Phys.\ Rev.\ D {\bf 44} (1991) 1100.

\bibitem{Mignemi:1991wa}
  S.~Mignemi and D.~L.~Wiltshire,
  Phys.\ Rev.\ D {\bf 46} (1992) 1475
  [hep-th/9202031].

\bibitem{Poletti:1994ff}
  S.~J.~Poletti and D.~L.~Wiltshire,
  Phys.\ Rev.\ D {\bf 50} (1994) 7260
   [Erratum-ibid.\ D {\bf 52} (1995) 3753]
  [gr-qc/9407021].

\bibitem{Mignemi:1999zy}
  S.~Mignemi,
  Phys.\ Rev.\ D {\bf 62} (2000) 024014
  [gr-qc/9910041].

\bibitem{Melis:2005xt}
  M.~Melis and S.~Mignemi,
  Class.\ Quant.\ Grav.\  {\bf 22} (2005) 3169
  [gr-qc/0501087].

\bibitem{Clifton:2005aj}
  T.~Clifton and J.~D.~Barrow,
  Phys.\ Rev.\ D {\bf 72} (2005) 103005
  [gr-qc/0509059].

\bibitem{standby}
A.~Ganguly, R.~Gannouji, R.~Goswami and S.~Ray,
  In~preparation.

\bibitem{Ellis:1966ta}
  G.~F.~R.~Ellis,
  J.\ Math.\ Phys.\  {\bf 8} (1967) 1171.

\bibitem{Stewart:1967tz}
  J.~M.~Stewart and G.~F.~R.~Ellis,
  J.\ Math.\ Phys.\  {\bf 9} (1968) 1072.

\bibitem{vanElst:1995eg}
  H.~van Elst and G.~F.~R.~Ellis,
  Class.\ Quant.\ Grav.\  {\bf 13} (1996) 1099
  [gr-qc/9510044].

\bibitem{Goswami:2011ft}
  R.~Goswami and G.~F.~R.~Ellis,
  Gen.\ Rel.\ Grav.\  {\bf 43} (2011) 2157
  [arXiv:1101.4520 [gr-qc]].

\bibitem{Roy:2014lda}
  X.~Roy,
  arXiv:1405.6319 [gr-qc].

\bibitem{Maeda1980}
K.~Maeda, M.~Sasaki, T.~Nakamura and S.~Miyama,
Prog.\ Theor.\ Phys.\  {\bf 63} (1980), 719-721.

\bibitem{Clarkson:2002jz}
  C.~A.~Clarkson and R.~K.~Barrett,
  Class.\ Quant.\ Grav.\  {\bf 20} (2003) 3855
  [gr-qc/0209051].

\bibitem{Clarkson:2007yp}
  C.~Clarkson,
  Phys.\ Rev.\ D {\bf 76} (2007) 104034
  [arXiv:0708.1398 [gr-qc]].

\bibitem{Betschart:2004uu}
  G.~Betschart and C.~A.~Clarkson,
  Class.\ Quant.\ Grav.\  {\bf 21} (2004) 5587
  [gr-qc/0404116].


\bibitem{Plebanski:2006sd}
  J.~Plebanski and A.~Krasinski,
 ``An introduction to general relativity and cosmology''(p. 162),
  Cambridge, UK: Univ. Pr. (2006) 534 p 

\bibitem{Sasaki1980}
M.~Sasaki, K.~Maeda, S.~Miyama and T.~Nakamura,
Prog.\ Theor.\ Phys.\  {\bf 63} (1980), 1051-1053.



\bibitem{Hamid:2014kza}
  A.~I.~M.~Hamid, R.~Goswami and S.~D.~Maharaj,
  Class.\ Quant.\ Grav.\  {\bf 31} (2014) 135010
  [arXiv:1402.4355 [gr-qc]].

\bibitem{Bousso:2002fq}
  R.~Bousso,
  hep-th/0205177.



\end{thebibliography}
\end{document}